%% file: masterfile.tex
\documentclass[10pt,twoside,BCOR7mm,DIV15,headinclude,footexclude,
               cleardoubleempty,idxtotoc]
{scrartcl}

\usepackage{natbib}
\usepackage[font=small,labelfont=bf]{caption}
\usepackage[english]{babel}
\usepackage{graphicx}
\usepackage{hyperref}
\usepackage{scrpage2}
\usepackage{ifthen}
\usepackage{booktabs}
\usepackage{amsmath}
\usepackage{amssymb}
\usepackage{multicol}
\usepackage{float}
\usepackage{hyperref}

\hypersetup{breaklinks=true
,colorlinks=true,linkcolor=black,urlcolor=blue
,citecolor=black}

\addto\captionsenglish{%
}

\makeatletter
\renewcommand{\@biblabel}[1]{}
\renewcommand{\@cite}[2]{%
{#1\ifthenelse{\boolean{@tempswa}}{,#2}{}}}
\makeatother
\ofoot{\thepage}
\ifoot{}

\setcapindent{0em}
\setheadsepline{1pt}

\setkomafont{pagehead}{\normalfont\sffamily}
\setkomafont{pagenumber}{\normalfont\rmfamily}

\makeatletter
\newcommand{\listofcontributions}{\@starttoc{con}}

\newcommand{\l@contribution} {\@dottedtocline{1}{1.5em}{2.3em}}
\makeatother

\newenvironment{contribution}{
\setcounter{section}{0}
\setcounter{figure}{0}
\setcounter{table}{0}
}{
\newpage
\lehead{}
\rohead{}
}

\begin{document}

\setlength{\baselineskip}{2.5ex}

\begin{contribution}
\include{myarticle}
\end{contribution}

%%-------------------------------------------------------

\end{document}

%% file: myarticle.tex
% EXAMPLE AND TEMPLATE FILE FOR PROCEEDINGS OF THE WOLF-RAYET WORKSHOP.
% PLEASE REPLACE THE TEMPLATE TEXT BY YOUR OWN ARTICLE.
% NOTE THAT YOU MUST NOT PROCESS THIS FILE, BUT THE MASTER FILE:
% latex masterfile; dvips masterfile

% RUNNING AUTHOR: PUT AUTHOR NAMED HERE
\lehead{B.\ Miszalski, R.\ Manick \& V.\ McBride}

% RUNNING TITLE; SHORTEN THE TITLE IF NECESSARY
% IN CASE OF A ONE-PAGE CONTRIBUTION (POSTER),
% SQUEEZE AUTHORS AND TITLE IN THIS LINE (Author: Title ...)
\rohead{Close binary Wolf-Rayet central stars of PNe}

\begin{center}
% FULL TITLE HEADING
{\LARGE \bf Post-common-envelope Wolf-Rayet central stars of planetary nebulae}\\
\medskip

% AUTHORS LIST
{\it\bf Brent Miszalski$^{1,2}$, Rajeev Manick$^3$ \& Vanessa McBride$^{1,4}$}\\

% AFFILIATIONS
{\it $^1$South African Astronomical Observatory, South Africa}\\
{\it $^2$Southern African Large Telescope Foundation, South Africa}\\
{\it $^3$KU Leuven, Belgium}\\
{\it $^4$University of Cape Town, South Africa}

% ABSTRACT
\begin{abstract}
   Nearly 50 post-common-envelope (post-CE) close binary central stars of planetary nebulae (CSPNe) are now known. Most contain either main sequence or white dwarf (WD) companions that orbit the WD primary in around 0.1-1.0 days. Only PN~G222.8$-$04.2 and NGC~5189 have post-CE CSPNe with a Wolf-Rayet star primary (denoted [WR]), the low-mass analogues of massive Wolf-Rayet stars. It is not well understood how H-deficient [WR] CSPNe form, even though they are relatively common, appearing in over 100 PNe. The discovery and characterisation of post-CE [WR] CSPNe is essential to determine whether proposed binary formation scenarios are feasible to explain this enigmatic class of stars. The existence of post-CE [WR] binaries alone suggests binary mergers are not necessarily a pathway to form [WR] stars. Here we give an overview of the initial results of a radial velocity monitoring programme of [WR] CSPNe to search for new binaries. We discuss the motivation for the survey and the associated strong selection effects. The mass functions determined for PN~G222.8$-$04.2 and NGC~5189, together with literature photometric variability data of other [WR] CSPNe, suggest that of the post-CE [WR] CSPNe yet to be found, most will have WD or subdwarf O/B-type companions in wider orbits than typical post-CE CSPNe (several days or months c.f. less than a day).
\end{abstract}
\end{center}

% TEXT OF THE PAPER, TWO-COLUMN STYLE
\begin{multicols}{2}

\section{Close binary central stars of Planetary Nebulae}
Planetary nebulae (PNe) are the ionised gaseous envelopes ejected by low to intermediate mass stars (1--8 $M_\odot$) at the end of the asymptotic giant branch (AGB) phase. The origin of the myriad shapes observed in PNe is a difficult problem that has persisted for several decades \citep{BalickFrank2002}. The most promising solution is now understood to be stellar or even sub-stellar companions \citep{DeMarco2009}. Observational evidence for binarity in PNe was relatively modest with pioneering photometric monitoring surveys producing around a dozen examples \citep{Bond2000}. Most of these have orbital periods of 0.1--1.0 d and have therefore passed through the common-envelope (CE) phase \citep{Ivanova2013}. As the ejected CE is still visible as the PN with an age of $\sim$10$^4$ years, post-CE central stars of planetary nebulae (CSPNe) are the youngest probes we have of the very short (unobserved) and poorly understood CE interaction. 

To realise the potential of post-CE CSPNe as a population to address problems concerning CE evolution, as well as the shaping mechanism of PNe, a statistically significant sample must be gathered. \cite{Misz2008, Misz2009a} leveraged the large areal coverage, high time cadence and high sensitivity of I-band photometry from the microlensing survey OGLE-III \citep{Udalski2008} to discover several more post-CE CSPNe. A major result from this work is that around 1 in 5 PNe have a post-CE CSPN, consistent with earlier estimates \citep{Bond2000}. High-quality imaging of the enlarged sample has revealed post-CE PNe often show collimated outflows or jets, low-ionisation filaments and rings, as well as bipolar morphologies \citep{Misz2009b}. These morphological features have since been successfully used to target PNe with similar morphologies to discover several more post-CE binaries \citep{MiszAPN5, Jones2015}, bringing the total to nearly 50 binaries \citep{Jones2015}.

The population of post-CE CSPNe mostly consists of white dwarf (WD) primaries with main sequence (MS) secondaries, however several double degenerate (DD) systems are also known \citep[e.g.][]{Tovmassian2010, BoffinMiszalski2012, Santander2015}. Photometric variability of WDMS CSPNe is often dominated by a large sinusoidal irradiation effect, produced by the heating of the hemisphere of the secondary facing the primary. Also present in post-CE CSPNe lightcurves are eclipses and ellipsoidal variability, the latter mostly occurring in DD systems. The optical photometric monitoring technique used in most surveys for post-CE CSPNe has introduced significant selection effects into the known population. Firstly, there is a strong bias against finding WDMS binaries with orbital periods $>$1 d since the irradiation effect amplitude decreases rapidly with increasing orbital separation \citep{DeMarcoHillwig2008}. Secondly, DD systems are under-represented since they (a) do not usually produce an irradiation effect, and (b) require a close orbital separation to produce ellipsoidal variability or eclipses. Radial velocity (RV) monitoring is therefore the only reliable technique to establish if a post-CE CSPN is present \citep[e.g.][]{BoffinMiszalski2012}. 

\section{Binary status of Wolf-Rayet central stars}
Over a hundred PNe have H-deficient Wolf-Rayet central stars (denoted [WR]), the low-mass ($M\sim0.6$ $M_\odot$) spectroscopic mimics of massive WR stars (see the review of \citet{Todt2015}, these proceedings). The origin of [WR] stars remains elusive, especially given the recent discovery of two bona-fide [WN] stars whose extremely He-rich atmospheres may be better explained by a merger scenario, rather than the traditional scenarios used to explain [WC] star formation \citep{Misz2012, Todt2013}. Some binary formation scenarios have been proposed \citep[e.g.][]{DeMarcoSoker2002}, but there are currently too few known binaries to further develop these scenarios. The only known [WR] binaries are PN~G222.8$-$04.2 \citep[$\lbrack$WC7$\rbrack$,][]{Hajduk2010} and NGC~5189 \citep[$\lbrack$WO1$\rbrack$,][]{Manick2015} with respective orbital periods of 1.26 and 4.05 d.

Why are there so few post-CE [WR] CSPNe known? The answer lies in the previous reliance on photometric monitoring to find post-CE CSPNe. At least 37 early-type [WC] stars and related H-deficient stars have been monitored photometrically by \citet{CiardulloBond1996} and \citet{Gonzalez2006}. Some were found to be non-radial pulsators, but no variability was found that could be attributed to a binary companion. Indeed, since 15--20\% of PNe have post-CE CSPNe \citep{Bond2000, Misz2009a}, then we would have expected $\sim$5--7 post-CE [WR] CSPNe to have been found. These non-detections exclude the presence of post-CE [WR] binaries in the sample with MS companions since they would have produced a very large amplitude irradiation effect as in e.g. \citet{Corradi2011} and \citet{ETHOS1}.

\section{A new approach: RV monitoring of [WR] stars}
RV monitoring is a tried and tested method to discover massive WR binaries \citep[e.g.][]{Foellmi2003}, so there is no reason not to suspect it would work for [WR] stars. Our approach is to take several spectra of each object with a resolution of $\sim$2\AA\ and signal-to-noise $>$ 40 in the continuum. We are using SpCCD on the SAAO 1.9-m telescope for the brightest targets and the Robert Stobie Spectrograph (RSS) on the Southern African Large Telescope (SALT) for fainter targets. We then follow the cross-correlation procedure of \citet{Foellmi2003} to derive the RV measurements. Initial results have been published in \citet{Manick2015}. No periodic variability was found in PMR~2, Hen~2-99, NGC~5315, Hen~2-113 and Hen~3-1333, whereas we found periodic variability of 4.04 d in NGC~5189. Additional higher quality SALT observations have since ruled out the previously suspected variability in Hen~2-99 and NGC~5315. 

\begin{figure*}[!ht]
\begin{center}
\includegraphics[scale=0.35]{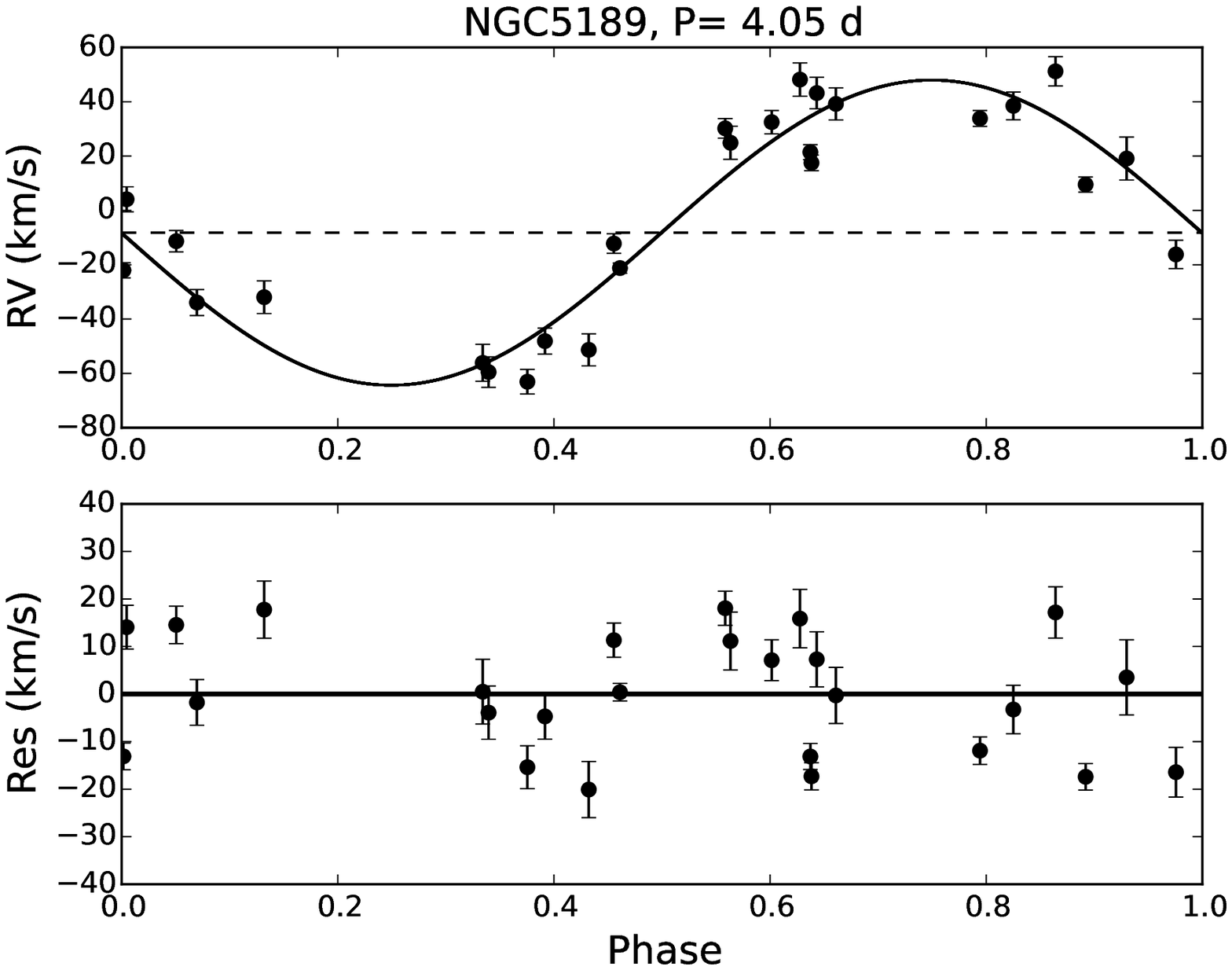}
 \includegraphics[scale=0.35]{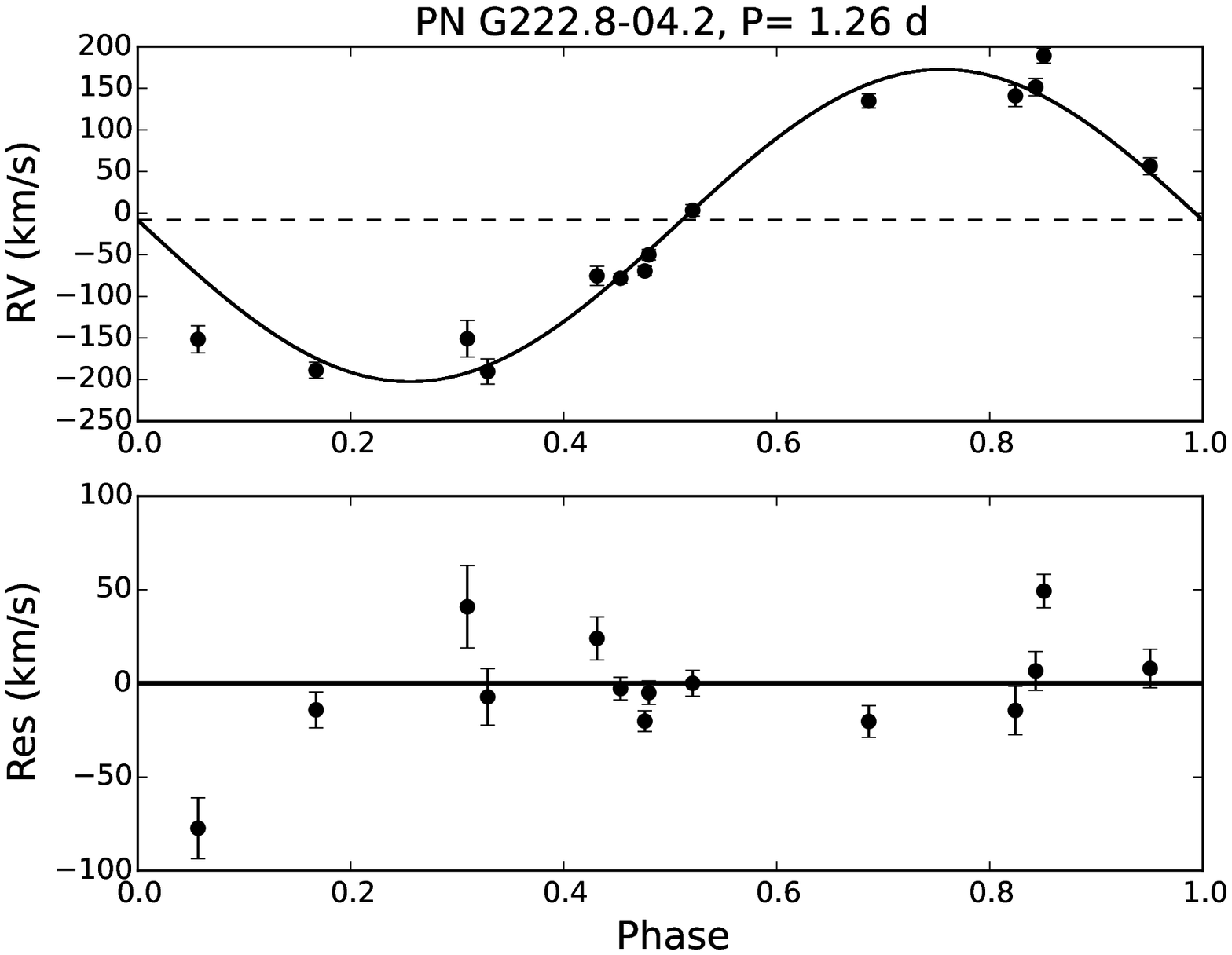}
 \caption{SALT RSS RV curves of NGC~5189 (left upper panel, 26 epochs) and PN~G222.8$-$04.2 (right upper panel, 14 epochs) fitted with circular orbits (solid curve). The residuals (lower panels) show a larger amplitude scatter than seen in non-[WR] post-CE CSPNe \citep[e.g.][]{BoffinMiszalski2012}, but is consistent with wind variability that is also encountered in massive WR RV curves \citep[e.g.][]{Foellmi2003}. The larger residuals in PN~G222.8$-$04.2 are probably the result of the cross-correlation method reacting to the high degree of line profile variability. 
\label{miszalski:rv}}
\end{center}
\end{figure*}

Figure \ref{miszalski:rv} presents an updated RV curve for NGC~5189 containing 12 additional spectra, resulting in a revised period of 4.05 d, as well as our preliminary RV curve for PN~G222.8$-$04.2. The measured RV amplitudes for NGC~5189 and PN~G222.8$-$04.2 are $54\pm4$ and $184\pm8$ km s$^{-1}$, respectively. As the orbital inclinations of both objects are unknown, we present a list of possible companion masses in Tab. \ref{miszalski:masses}. An evolved WD companion is most likely present in NGC~5189, similar to that in Fleming~1 \citep{BoffinMiszalski2012}. In the case of PN~G222.8$-$04.2, if the unusually high RV amplitude is not an artefact of the cross-correlation technique, then we may suspect the companion to be a subdwarf O/B-type star, perhaps similar to that found in NGC~6026 \citep{Hillwig2010}. Polarimetric observations and spatiokinematic modelling \citep[e.g.][]{Sabin2012} would be desirable to constrain the inclinations of both objects. 

\begin{table}[H] 
\begin{center} 
   \captionabove{Companion mass estimates for a variety of orbital inclination angles assuming $M_1=0.6$ $M_\odot$.}
\label{miszalski:masses}
\begin{tabular}{lcc}
\toprule
$i$ ($^\circ$) & $M_2$ (NGC~5189) & $M_2$ (PN~G222.8$-$04.2)\\
\midrule 
30 & 1.30$\pm$0.20  & 7.60$\pm$0.70\\
40 & 0.80$\pm$0.10  & 4.10$\pm$0.20\\
50 & 0.59$\pm$0.04  & 2.70$\pm$0.10\\
60 & 0.50$\pm$0.02  & 2.10$\pm$0.10\\
70 & 0.44$\pm$0.01  & 1.80$\pm$0.10\\
80 & 0.42$\pm$0.01  & 1.60$\pm$0.03\\
90 & 0.40$\pm$0.01  & 1.60$\pm$0.04\\
\bottomrule
\end{tabular}
\end{center}
\end{table}

\section{Summary}
We have started to conduct RV monitoring of several [WR] CSPNe to search for post-CE [WR] binaries. These binaries are needed to provide the first accurate distance-independent stellar parameters for [WR] stars, as well as to inform the still enigmatic formation scenarios regarding [WR] stars. We have increased the known sample of post-CE [WR] binaries to two with the discovery of a 4.05 d orbital period in NGC~5189 \citep{Manick2015}, an object long suspected of harbouring a binary CSPN \citep{PhillipsReay1983} because of its peculiar morphology \citep{Sabin2012}. Its morphology is typical of post-CE PNe and in particular bears close resemblance to the post-CE PN NGC~6326 \citep{NGC6326}. Furthermore, we have presented a preliminary RV curve for PN~G222.8$-$04.2 that will form the basis of a more detailed study to be presented elsewhere. 

The anomalously large 4.05 d orbital period for NGC~5189 suggests that the orbital periods of other post-CE [WR] CSPNe may also be several days or months. This may be a result of the [WR] primary and its extended atmosphere requiring more space in its orbit. Such binaries can only be found from RV monitoring and much work remains to find them. If this is correct, then we would expect close binaries containing the progeny of [WC] stars, PG1159 stars, to have similarly wide orbits. 

%\section{Acknowledgements}
This work incorporates SALT observations taken under programmes 2013-2-RSA-005 and 2014-2-SCI-060.
We thank Tony Moffat for helpful discussions.

\bibliographystyle{aa}
\bibliography{myarticle}

\end{multicols}